\documentstyle[preprint,aps]{revtex}
\begin{document}

\newcommand\del{\mbox{\boldmath $\nabla$}}
\newcommand\bfv{\text{\bf v}}

\tighten

\title {PECULIAR VELOCITY, COSMIC PERTURBATION \\
THEORY AND THE CMB ANISOTROPY}

\author{{\sc Marco Bruni}
\thanks{email bruni@uk.ac.qmw.star}
}
\address{School of Mathematical Sciences, Queen Mary and Westfield
College, \\ University of London, Mile End Road, London E1 4NS, UK.\\
Dipartimento di Astronomia, Universit\`a degli Studi di Trieste,\\
 via Tiepolo 11, 34131 Trieste, Italy.
}
\author{{\sc David H. Lyth}
\thanks{email lyth@uk.ac.lancs.ph.v1}
}
\address{School of Physics and Materials, Lancaster University,
Lancaster LA1 4YB, UK. \\ $\mbox{}$ }

\date{July 1993, this extended version November 1993}
\thispagestyle{empty}
\maketitle

\begin{abstract}

It is proved that the cosmological density perturbation is associated
with a peculiar velocity field. This allows a simple formulation of
cosmological perturbation theory, which works entirely with
quasi-Newtonian fluid flow equations.  As an illustration, the large
scale cosmic microwave background anisotropy (Sachs-Wolfe effect) is
calculated without any reference to the metric perturbation.
In addition, assuming the usual adiabatic initial condition  on the
density perturbation, we show that the dipole of the anisotropy
 measures our peculiar velocity
relative to the average peculiar velocity within the last scattering
surface of the microwave background, thus defining its frame of
reference.

\end{abstract}



\newpage

The interplay between particle physics and cosmology,
generally known as particle cosmology, is
particularly fruitful at the present time, and is likely to
play a decisive role in both fields during the next few years.
For some time, the main focus has has been on the
origin of structure in the universe. In this context,
one is seeking a huge body of data on
galaxies and galaxy clusters,
as well as the cosmic microwave background (cmb) anisotropy.
The
first observation \cite{smet} in 1992 of the intrinsic part of this isotropy
has been one reason for
for a dramatic surge of interest in structure
formation, and the ever improving quality of data
on galaxies and clusters has been another.

The simplest hypothesis concerning the origin of large scale structure
is to invoke a primeval density perturbation (plus perhaps a
gravitational wave contribution to the cmb anisotropy), which originates
during inflation as a vacuum fluctuation. If this hypothesis is correct,
the observations mentioned above provide a window on the nature of the
fundamental interactions at an energy scale which is many orders of
magnitude bigger than those which can be explored at colliders.

In order to explore the hypothesis, one needs to formulate
a relativistic theory of cosmological perturbations.
At present three rival formulations exist in the literature. The
`metric perturbation' formulation, originated by Lifshitz in 1946
\cite{lifs}, considers the components of the metric tensor, which are
related to the components of the energy-momentum tensor in either the
perfect fluid approximation, or including the effects of particle
diffusion and free-streaming \cite{peeb}.  The equations in this
approach are  complicated, though they are perfectly serviceable and
have in fact been used in most of the decisive comparisons of theory
with observation.

Second, there is the `gauge invariant' approach  initiated by Bardeen
in 1980 \cite{bard} and extended by various authors to include diffusion
and free-streaming \cite{kosa}. It
starts with the metric perturbation
approach, but allows one to derive a much simpler set of equations for
 matter perturbation variables   which do not
explicitly involve the metric perturbation. The only problem is that
the physical interpretation of the variables used is somewhat
obscure, as is that of the mathematical manipulations through which the
equations are derived.

 Finally
there is an approach pioneered by
Hawking\cite{hawk} in 1966 but developed mostly in the last decade
\cite{lymu,elbr,lyst,bret,duet,lily}
which is
based on the `covariant' fluid flow approach to general
relativity\cite{bi:ellis}. Here, one bypasses the metric tensor to
work directly with the curvature tensor, which allows a simple
treatment of perturbations, even permitting a coordinate-free, gauge invariant
description\cite{elbr,bret,duet}. This approach has been
developed for a single perfect fluid \cite{lymu,elbr,lyst,bret} and for
several uncoupled perfect fluids \cite{lyst,duet}, and in both cases
it has been shown to provide a simple, physically transparent,
derivation of the main results of the `gauge invariant' approach.
 Although
all of the `gauge invariant' variables have been recently
identified with physical quantities\cite{bret,duet}, no relation has
yet been provided with one of the most important observables, i.e.
with the peculiar velocity field; in addition, diffusion and free
streaming remain to be treated.

Something which has hitherto  generally  being lacking in
relativistic perturbation theory  is a straightforward procedure
for going to the Newtonian limit.  Moreover,  the concept of
peculiar velocity, which plays a vital role in the Newtonian case, has
not so far  even  been mentioned in the relativistic context.

In this letter we show, using the `covariant' approach,
how to extend the peculiar velocity (PV) concept to the
relativistic regime. As an
application of the concept, we clarify the relation between the PV field
and the
cosmic microwave background (CMB) anisotropy, deriving the Sachs Wolfe
effect in a straightforward way, without any mention to metric
perturbations. Furthermore, we show that the dipole of the anisotropy
measures our PV with respect to the average PV within the last
scattering surface of the CMB, thus defining for the first time a
cosmic frame of reference. We end by relating our peculiar velocity
to one of the `gauge invariant' variables, and pointing to the wider
implications of our work.

This letter is necessarily brief with emphasis on the Newtonian
analogy: full results, as well as a clarification of the link between
the treatment here and the other possible approaches mentioned before,
will be reported elsewhere \cite{brly}.

The universe is regarded
as a fluid, taken here to be perfect, and a comoving
observer measures by definition zero momentum density, energy density
$\rho$ and pressure $p$. At a given spacetime point one can use
locally inertial coordinates $y^\mu$ in which the fluid is
instantaneously at rest.
Then  \cite{1} the fluid 4-velocity $u^\mu$ at (infinitesimally) nearby
spacetime points
defines the acceleration $a_i=\partial u_i/\partial y^0$ of the fluid,
and its velocity gradient $u_{ij}\equiv \partial u_j/\partial y^i$.
The velocity gradient can be uniquely decomposed into an antisymmetric
vorticity $\omega_{ij}$, a symmetric traceless shear $\sigma_{ij}$,
and a locally defined Hubble parameter $H$,
 \begin{equation}
u_{ij}=H \delta_{ij}+\sigma_{ij} +\omega_{ij}\;. \label{14}
 \end{equation}
The unperturbed universe is isotropic around each comoving observer,
in particular the acceleration, shear and vorticity all vanish. Since
the vorticity vanishes there exist hypersurfaces of simultaneity,
orthogonal to the fluid worldlines. Isotropy
about every comoving observer implies that these hypersurfaces are
homogeneous; on them $\rho$, $p$ and $H$ are position-independent, and
so is the proper time $t$ along a fluid worldline, starting from one of
the hypersurfaces.
Making the assumption of critical density $ H^2=
(8\pi G/3)\rho $,
the hypersurfaces are flat and one can
define on them coordinates $x_i$ which are fixed along each fluid
worldline (comoving), such that $r_i\equiv a x_i$ are Cartesian coordinates
(the generalisation to non-critical density is straightforward
\cite{brly}). The scale factor, defined by $aH=da/dt$, is normalised to
1 at the present epoch.

Newtonian physics is valid after matter domination, on scales much smaller
than the Hubble distance $H^{-1}$. In this case there is a
well defined fluid velocity ${\text{\bf u}}$, and a PV field
${\text{\bf v}}$ related to it by
\begin{equation}
{\text{\bf u}}({\text{\bf r}})-{\text{\bf u}}(0)= \bar H {\text{\bf r}}+
{\text{\bf v}}(
{\text{\bf r}})-{\text{\bf v}}(0)\;,
\label{9}
\end{equation}
where ${\text{\bf r}}$ is the displacement from our position
(or that of any other
comoving observer), and $\bar{H}$ the average expansion.
An equivalent statement in terms of the velocity gradient is
\begin{equation}
u_{ij}=\delta_{ij}\bar H+\partial_i v_j\;.
\label{9a}
\end{equation}
These expressions define ${\text{\bf v}}$ up to a constant, which can be
chosen so that the average of ${\text{\bf v}}$ vanishes. Making the
decomposition \cite{2}
${\text{\bf v}}={\text{\bf v}}^{\text L}+{\text{\bf v}}^{\text T}$, where
the transverse part ${\text{\bf v}}^{\text T}$
satisfies $\partial_i v^{\text T}_i=0$
and the longitudinal part is of the form
 ${\text{\bf v}}^{\text L}={\mbox{\boldmath $\nabla$}}
\psi_v$,
the  comparison with (\ref{14}) gives
 \begin{eqnarray}
\omega_{ij} &=& \frac12 (\partial_i v^{\text T}_j-\partial_j v^{\text T}_i)\;,
\label{21} \\
\delta H &=& \frac13 \nabla^2 \psi_v\;,
\label{23}
\end{eqnarray}
where $\delta H$ is the perturbation in the expansion,
while the shear $\sigma_{ij}$ does not carry any new information.
Defining the peculiar gravitational potential $\psi$ by
\begin{equation}
4\pi G\delta \rho = \nabla^2\psi\;,
\label{psi}
\end{equation}
the Newtonian hydrodynamical equations show that it is constant
(dropping the decaying mode), and that
\begin{equation}
{\text{\bf v}}^{\text L}({\text{\bf x}},t)
=-t {\mbox{\boldmath $\nabla$}}\psi({\text{\bf x}})\;.
\label{vl}
\end{equation}

Now consider the relativistic regime.
In the presence of perturbations, the special coordinates $t$
and $x_i$ do not exist, but any first order approximation to them
may be used to describe the perturbations, with an error of only second
order. {\em First order perturbations `live' in the unperturbed universe.}
We use the notation $\partial_i\equiv\partial/\partial r_i
=a^{-1} \partial/\partial x_i$, and let an overdot denote
differentiation with respect to t at fixed ${\text{\bf x}}$.
Writing each perturbation $f({\text{\bf x}},t)$ as a
Fourier series in a comoving box much bigger than the region of
interest, modes with different
wave-vector ${\text{\bf k}}/a$
decouple in the equations.
A mode is said to
`enter  the horizon' when $aH/k$ falls below 1,
and for
scales
$k^{-1}\mathrel{\rlap{\lower4pt\hbox{\hskip1pt$\sim$}}
    \raise1pt\hbox{$>$}}
200{\,\mbox{Mpc}}$ this occurs after matter-domination.

Because they are time dependent,
perturbations in $\rho$, $p$ and $H$ have to be
defined with respect to a slicing of spacetime into
spacelike hypersurfaces \cite{8},
which become homogeneous in the unperturbed limit.
On each hypersurface we split $\rho$ into an average plus a
perturbation, $\rho=\bar\rho+\delta\rho$, and similarly for $p$,
$H$.

We are now going to show that the slicing can be chosen so that
there exists a relativistic generalization of
Eq.~(\ref{9a}) which
defines a unique longitudinal PV ${\text{\bf v}}^{\text L}
={\mbox{\boldmath $\nabla$}}\psi_v$.
This relativistic generalization of the Newtonian
concept of PV is the central result of this letter.

If the vorticity
does not vanish there are no hypersurfaces orthogonal to the fluid
worldlines, but Eq.~(\ref{21}) defines
\cite{2,7} a unique ${\text{\bf v}}^{\text T}$. Then we
can define modified fluid
worldlines which at each spacetime point have velocity
 $-{\text{\bf v}}^{\text T}$ with
respect to a comoving observer.  These worldlines
have zero vorticity, so there exist
hypersurfaces orthogonal to them, called comoving
hypersurfaces\cite{footngauge},
on which we define $\delta \rho$, $\delta p$ and $\delta H$.

Equations involving only
$\omega_{ij}$, $\sigma_{ij}$,
$\delta\rho$, $\delta p$ and $\delta H$
can be derived \cite{hawk,lymu,elbr}
from the Einstein field equation, together with the Ricci identity
acting on
the 4-velocity $u^\mu$.
One of them is the constraint equation
\begin{equation}
2\partial_i \delta H + \dot H v^{{\text T}}_{\,i} =
\partial_j (\sigma_{ij}+\omega_{ij})\;,
\label{200}
\end{equation}
(The left hand side is the derivative perpendicular to the fluid
worldlines, whereas the first term alone is the derivative perpendicular
to the modified worldlines, i.e., within the comoving hypersurfaces).
As we now show, this equation is equivalent to Eq.~(\ref{9a}) in the Newtonian
limit, and gives the desired generalisation of it in the relativistic
regime.
We start with
the fact that any traceless symmetric tensor field may be decomposed
uniquely \cite{2}
in the form
\begin{equation}
 \sigma_{ij}=\frac12(\partial_i w^{\text T}_j+\partial_j w^{\text T}_i)
+\nabla_{ij}\chi_v+\sigma^{\text T}_{ij}\;, \label{split}
\end{equation}
where
$\nabla_{ij}\equiv \partial_i\partial_j-\frac13 \delta_{ij}
\nabla^2$, and
${\text{\bf w}}^{\text T}$ and $\sigma^{\text T}_{ij}$ are
transverse, $\partial_i
w
^{\text T}_i=0$ and $\partial_i \sigma^{\text T}_{ij}=0$.
Also, any scalar field $\delta H$ may \cite{2} be written
uniquely in the form
Eq.~(\ref{23}).
Then, using
Eqs.~(\ref{21}), (\ref{23}) and (\ref{split}),
the longitudinal part of Eq.~(\ref{200})
is $\chi_v=\psi_v$, and its
transverse part is
in Fourier space
\begin{equation}
{\text{\bf w}}^{\text T}= \left[1+6\left(1+\frac{p}{\rho} \right)
\left(\frac{aH}{k}\right)^2 \right] {\text{\bf v}}^{\text T}\;.\label{vrel}
\label{wt}
\end{equation}
 Calculating $\delta u_{ij}=\delta H\delta_{ij}+\sigma_{ij}+\omega_{ij}$
one therefore finds
\begin{eqnarray}
\delta u_{ij} &=& \partial_i  v^{\text L}_j +\sigma_{ij}^{\text T}+
\nonumber\\
&&\frac12 (\partial_i v^{\text T}_j - \partial_j v^{\text T}_i)+
\frac12 (\partial_i w^{\text T}_j + \partial_j w^{\text T}_i)\;.
\label{duij}
\end{eqnarray}

Thus, in the relativistic regime there is a well defined longitudinal
PV ${\text{\bf v}}^{\text L}$, but
two different `transverse peculiar
velocities' ${\text{\bf v}}^{\text T}$ and ${\text{\bf w}}^{\text T}$,
 related by Eq.~(\ref{vrel}).
 The difference
is a purely relativistic effect (the
dragging of inertial frames)
and one verifies from Eq.~(\ref{vrel}) that it disappears
in the Newtonian regime
$aH/k\ll1$.

The symmetric, traceless, transverse
term $\sigma_{ij}^{\text T}$  directly represents the effect on matter motion
of gravitational waves, a degree of freedom absent in Newtonian
physics.

The other equations involving only $\omega_{ij}$, $\sigma_{ij}$,
$\delta\rho$, $\delta p$ and $\delta H$ are propagation equations, that
in the Newtonian limit are equivalent to the usual fluid flow
equations. One of them
\cite{bret}
involves only the vorticity $\omega_{ij}$, and shows that it
decays like $(\rho+p)^{-1} a^{-5}$,
which presumably means that it is negligible.
The others are
\cite{lymu},
\begin{eqnarray}
(\delta\rho)\dot{} &=&-3(\rho+p)\delta H -3H\delta\rho\;,
\label{12} \\
(\delta H)\dot{} &=&-2H\delta H-\frac{4\pi G}{3}\delta \rho -\frac13
\frac{\nabla^2 \delta p}{\rho+p}\;.
\label{112}
\end{eqnarray}
After matter domination
they lead to
the Newtonian  expressions
Eqs.~(\ref{psi}) and (\ref{vl}), even on scales
$aH/k\gg1$
where Newtonian physics does not apply.

The constraint equation and the three evolution equations
completely determine the evolution of the velocity gradient, except for the
gravitational wave contribution $\sigma_{ij}^{\text T}$.
This is related to the traceless transverse part $h^{\text{TT}}_{ij}$
of the spatial metric perturbation by
$\sigma^T_{ij}=\frac 12 \dot h^{\text{TT}}_{ij}$ \cite{bret}, which in turn
satisfies \cite{star}
\begin{equation} \ddot h^{\text{TT}}_{ij}+ 3 H \dot h^{\text{TT}}_{ij}-
\nabla^2 h^{\text{TT}}_{ij}=0\;.
\end{equation}
Well before horizon entry,
each Fourier component
$h^{\text{TT}}_{ij}(t,\text{\bf k})$ has
some constant value $A_{ij}(\text{\bf k})$
(ignoring a decaying mode). Well afterwards, it oscillates as a
standing wave with amplitude decreasing like $a^{-1}$,
and wavenumber and angular frequency $k/a$. For the scales
$k^{-1}
\mathrel{\rlap{\lower4pt\hbox{\hskip1pt$\sim$}}
    \raise1pt\hbox{$>$}}
\,100\mbox{Mpc}$ which enter the horizon after
matter domination (when $a^{-1}\propto (aH)^2$)
its contribution to the
velocity gradient is therefore $\delta u
_{ij}(\mbox{grav})\sim A_{ij} H^2 a/k $.

Let us ask how significant this contribution is
compared with the density perturbation
contribution $\delta u_{ij}(\mbox{dens})\sim \partial_i v_j^{\text L}$.
{}From Eq.~(\ref{vl}),
\begin{equation} \frac{\delta u_{ij}(\text{grav})}
{\delta u_{ij}(\text{dens})}\sim
\left(\frac{aH}{k}\right)^3
\frac{A_{ij}}{\psi}\;.
\label{grde}
\end{equation}
In general there is no reason why the ratio should not be large
at horizon entry, in which case the gravitational waves could still be
important well after horizon entry. However, if
$\psi$ and $A_{ij}$ both originate as a vacuum fluctuation during
inflation, then \cite{lily,star,infl} $A_{ij}
\mathrel{\rlap{\lower4pt\hbox{\hskip1pt$\sim$}}
    \raise1pt\hbox{$<$}}\psi$
(on the scales $k^{-1}
\mathrel{\rlap{\lower4pt\hbox{\hskip1pt$\sim$}}
    \raise1pt\hbox{$>$}}
100\,\mbox{Mpc}$ that we are considering.)
 This implies that
on the scale $k^{-1}\sim 100\,\mbox{Mpc}$,
$\delta u_{ij}(\text{grav})$
is $\mathrel{\rlap{\lower4pt\hbox{\hskip1pt$\sim$}}
    \raise1pt\hbox{$<$}}10^{-6}\delta u_{ij}(\text{dens})$
at the present epoch, amply justifying its neglect when
$\text{\bf v}^{\text L}$ is deduced from large scale galaxy surveys.

To demonstrate the utility of Eq.~(\ref{duij}),
we now use it
\cite{lily} to calculate the Sachs-Wolfe
effect \cite{sawo},
which is normally written in terms of the metric perturbation,
and describes part of the anisotropy of the CMB.
This anisotropy is defined as
the angular variation in the intensity of the background at fixed
wavelength $\lambda$,
and is usually specified by giving the equivalent variation
$\Delta({\text{\bf e}})\equiv \Delta T/T$ in the temperature
$T$ of the blackbody distribution, where ${\text{\bf e}}$ is a unit vector
pointing in the direction of observation.
Some anisotropy $\Delta_{\text{em}}({\text{\bf e}})$
is already present on a comoving hypersurface just after
last scattering, and the Sachs-Wolfe effect describes the additional
anisotropy acquired on the journey towards us, to first order in
the perturbations. The Sachs-Wolfe effect
is due entirely to the anisotropy of the redshift
of the CMB, as the photons path can be taken as
unperturbed in first order.

Consider a photon passing a succession of comoving observers.
 Its trajectory is $d {\text{\bf r}}/dt=-{\text{\bf e}}$ and
between nearby observers its Doppler shift is
\begin{equation}
-\frac{d\lambda}{\lambda}=e_i e_j u_{ij} dr=-\frac
{d\bar a}{\bar a} + e_i e_j \delta u_{ij} dr\;,
\end{equation}
where the first term is due to the average expansion, and the second
is due to the relative PV of the observers.
Integrating this expression gives
the redshift of radiation received by us, which was emitted from
a distant comoving source. The unperturbed result is
$\lambda/\lambda_{\text{em}}=1/a_{\text{em}}$, and the first order perturbation
gives
the Sachs-Wolfe effect
\begin{equation}
\Delta ({\text{\bf e}})-\Delta_{\text{em}}({\text{\bf e}})
=\int^{x_{\text{em}}}_0 e_i e_j \delta u_{ij}({\text{\bf x}},t) a(t) dx\;,
\label{delt}
\end{equation}
where $x_{\text{em}}\simeq 2H_0^{-1}$ is the coordinate distance of the
last scattering surface, and the integration is along the
photon trajectory
\begin{equation}
x(t)=\int^{t_0}_t \frac{dt}{a} = 3
\left( \frac {t_0}{a_0}- \frac ta \right) \label{xtee}\;.
 \end{equation}

The symmetric part of Eq.~(\ref{duij}) gives the various contributions
to the Sachs-Wolfe effect. The transverse PV is presumably
absent, and we have nothing new to say about the gravitational wave
contribution. This leaves ${\text{\bf v}}^{\text L}$, which after
integrating by parts and using Eqs.~(\ref{xtee}) and (\ref{vl}) gives

\begin{eqnarray}
\Delta ({\text{\bf e}})&=&
\Delta_{\text{em}}({\text{\bf e}})+
\frac 13 [\psi({\text{\bf x}}_{\text{em}} )-\psi( 0)]+
\nonumber\\
&&{\text{\bf e}}\cdot[{\text{\bf v}}(0,t_0)
-{\text{\bf v}}({\text{\bf x}}_{\text{em}},t_{\text{em}})]\;.
\label{2}
\end{eqnarray}
Here ${\text{\bf x}}_{\text{em}}=x_{\text{em}} {\text{\bf e}}$, and for clarity
${\text{\bf v}}^{\text L}$ is denoted by ${\text{\bf v}}$.

A better expression follows if one uses the divergence theorem and
Eq.~(\ref{xtee}) to project out the dipole part of
$\psi({\text{\bf x}}_{\text{em}})/3$.
One finds that it is equal to
$\langle {\text{\bf v}}(t_{\text{em}})- {\text{\bf v}}(t_0) \rangle$
where $\langle\rangle$ denotes the average within the last scattering surface
$x=x_{\text{em}}$. Defining $\tilde{\text{\bf v}}=
{\text{\bf v}}-\langle{\text{\bf v}}
\rangle$
this gives
\begin{eqnarray}
\Delta ({\text{\bf e}})&=& \Delta_{\text{em}} ({\text{\bf e}})+
\frac 13 \left[\psi({\text{\bf x}}_{\text{em}} )\right]_{l>1}+
\nonumber\\
&&{\text{\bf e}}\cdot\left[ \tilde{\text{\bf v}}(0,t_0)-
\tilde{\text{\bf v}}({\text{\bf x}}_{\text{em}},t_{\text{em}})
\right]\;.
\label{best}
\end{eqnarray}

On angular scales $\mathrel{\rlap{\lower4pt\hbox{\hskip1pt$\sim$}}
    \raise1pt\hbox{$>$}}1^0$, corresponding to linear scales
at last scattering which are outside the horizon,
$\tilde {\text{\bf v}}({\text{\bf x}}_{\text{em}},t_{\text{em}})$
is negligible compared with $
\psi$, and with the usual adiabatic initial condition
so is $\Delta_{\text{em}}({\text{\bf e}})\simeq\frac43\delta
\rho({\text{\bf x}}_{\text{em}},t_{\text{em}})
/\rho(t_{\text{em}})$ \cite{lily,peeb}.
Dropping the monopole one then obtains
\begin{equation}
\Delta ({\text{\bf e}})=\frac13 \left[\psi({\text{\bf x}}_{\text{em}} )
\right]_{l>1} + {\text{\bf e}}\cdot\tilde{\text{\bf v}}_0\;.
\label{424}
\end{equation}

The last term in this expression is the dipole, and defines for the
first time the rest frame of the CMB (in
linear theory with the adiabatic initial condition). This is simply the
frame at rest with respect to the average peculiar velocity of
everything within the last scattering surface. Of course this result
corresponds {\em roughly} to the intuitive idea that a distant density
perturbation will affect the matter and the CMB
almost equally \cite{grze}.

Although the above application of the peculiar velocity concept is
amusing, it is far from being the most important aspect of our work
and we end by briefly summarizing the wider picture,
dropping for  simplicity the rotational and gravitational wave modes.

Here we have been using the `covariant' \cite{lily} fluid flow
approach to cosmological  perturbation theory as opposed to the
`metric'\cite{peeb} and `gauge invariant' \cite{bard} approaches,
showing how close is our treatment to the Newtonian one. The link
between the two latter has been deeply analyzed\cite{kosa}, while the
relations between the main variables of the  `covariant' approach  and
those of the `gauge invariant' treatment were given\cite{bret,duet},
but not yet fully exploited to treat diffusion and free--streaming.

Our new point of contact with the `gauge invariant' approach is the
fact that the Bardeen\cite{bard}  variable $v^{(0)}_s$ is related to
our peculiar velocity potential $\psi_v$. In fact,  Bardeen pointed out
in that original reference (Eqs.~(3.11) plus the one before
Eq.~(3.12)) that the shear of the velocity field
associated with $v_s^{(0)}$ is just the shear of the
comoving worldlines.
What we
have shown here is that, in a comoving slicing\cite{footngauge}
and a critical density (flat) universe, there is a unique quantity
$\bfv^{\text L}=\del \psi_v$, which gives both the shear of the
comoving worldlines {\it and} the perturbation in their expansion
through the relation $3\delta H=\del.\bfv$.
It follows that, under these conditions,
the velocity field associated with
$v_s^{(0)}$ is precisely the peculiar velocity defined in the present
paper.

 Adding this connection to the already known one\cite{bret,duet}, it
should now be possible to write {\it and simply derive} all of the
equations of the gauge invariant approach following a covariant
treatment. The result will be  a complete description of cosmological
perturbation theory, simpler than either of the alternatives and
treating the Newtonian and relativistic regimes on essentially the
same footing.

While this manuscript was in preparation, we became aware that a
related derivation of the Sachs-Wolfe effect was under
study\cite{rsxd}. We are grateful to P. Coles, F. Lucchin, R.
Scaramella for useful comments. MB thanks SERC (UK) and MURST (Italy)
for financial support.


\begin{references}
\bibitem{smet} G. F. Smoot {\it et. al.}, Astrophys. J. Letts.
{\bf 396}, L1 (1992).
\bibitem{lifs}
E. M. Lifshitz J. Phys. (Moscow) {\bf 10}, 116 (1946).
\bibitem{peeb} P. J. E. Peebles, {\sl Principles of Physical
Cosmology}, (Princeton University Press, N.J., 1993);
G. Efstathiou, in {\sl The Physics of the Early Universe},
eds. A.	Heavens, J. Peacock A. Davies (SUSSP publications, Edinburgh,
1990).
\bibitem{bard}
J. M. Bardeen, Phys. Rev. D {\bf 22}, 1882 (1980).
V.F. Mukhanov, H. A. Feldman and R. H. Brandenberger,
Phys. Rep. {\bf 215}, 203 (1992).
\bibitem{kosa}
H. Kodama and M. Sasaki, Prog. Theor. Phys., {\bf 78}, 1 (1984),
and references traced from N. Sugiyama, N. Gouda and M.
Sasaki, Astrophys. J. {\bf 365}, 432 (1990).
R. K. Schaefer, Int. J. Mod. Phys. {\bf A6}, 2075 (1991).
R. K. Schaefer and Q. Shafi, Phys. Rev. D {\bf 47}, 1333 (1993).
\bibitem{hawk}
Hawking, S. W.,  Astrophys. J., {\bf 145}, 544 (1966).
A. K. Raychaudhuri, {\sl Theoretical Cosmology}
(Clarendon Press, Oxford, 1979).
D. W. Olson, Phys. Rev. D, {\bf 14}, 327 (1976).
\bibitem{lymu} D. H. Lyth and M. Mukherjee, Phys. Rev. D {\bf 38}, 485 (1988).
\bibitem{elbr}
G. F. R. Ellis and M. Bruni, Phys. Rev. D,
{\bf  40}, 1804 (1989).
\bibitem{lyst}
D. H. Lyth and  E. D. Stewart, Astrophys. J., {\bf 361}, 343 (1990).
\bibitem{bret}
M. Bruni, P. K. S. Dunsby and G. F. R. Ellis,
Astrophys. J. {\bf 395}, 34 (1992),
and other references cited there.
\bibitem{duet}
P. K. S. Dunsby, M. Bruni and G. F. R. Ellis,
Astrophys. J. {\bf 395}, 54 (1992).
\bibitem{lily} A. R. Liddle and D. H. Lyth, Phys. Rep. {\bf 231}, 1
(1993).
\bibitem{bi:ellis}
G. F. R. Ellis, in {\sl Proceedings of XLVII Enrico Fermi Summer School},
ed R. K.  Sachs (Academic Press, 1971).
\bibitem{brly} M. Bruni and D. H. Lyth, in preparation.
\bibitem{1} Letters $i$, $j$, \dots denote space indices, and
locally we take
the line element to be $ds^2=-(dy^0)^2+\sum_i (dy^i)^2$, so that
there is no distinction between upper and lower space indices.
\bibitem{2}
This is easily proved for a given Fourier component,
taking ${\bf k}$ along say the $3$ axis. See also
J.M. Stewart, Class. Quantum Grav. {\bf 7}, 1169 (1990).
\bibitem{8}
This slicing can be loosely referred to as a gauge choice
\cite{kosa}; the gauge problem is
discussed in Ref. \cite{bard,elbr}. Quantities that
vanish in the isotropic universe
like $\sigma_{ij}$ and $\omega_{ij}$ are gauge invariant.
\bibitem{7}
This statement is equivalent to the equation $\epsilon_{ijk}\partial_i
\omega_{jk}=0$, derived in the covariant approach from the symmetry
properties of the curvature tensor.
\bibitem{footngauge}
Sometime this slicing is also called `velocity orthogonal slicing', as
e.g. in \cite{kosa}, section III(2).
\bibitem{star}
This equation is easily derived by perturbing the metric \cite{peeb}.
At the classical level one can dispense with the metric perturbation,
and work entirely with an equivalent differential equation for
$\sigma_{ij}^{\text T}$ which can be derived within the covariant approach.
However, according to present ideas cosmological gravitational waves
originate as a quantum fluctuation during inflation, and
for the quantum theory one needs the metric perturbation; see
A. A.  Starobinsky,  Sov. Astron. Lett. {\bf 11}, 133 (1985).
\bibitem{infl}
If the spectra of $A_{ij}$ and $\psi$ are fairly flat, the same thing
follows by comparing galaxy surveys (giving $\psi$ for
$10\,\mbox{Mpc}
\mathrel{\rlap{\lower4pt\hbox{\hskip1pt$\sim$}}
    \raise1pt\hbox{$<$}}k^{-1}
\mathrel{\rlap{\lower4pt\hbox{\hskip1pt$\sim$}}
    \raise1pt\hbox{$<$}}100\,\mbox{Mpc}$)
with the cmb anisotropy (bounding $A_{ij}$ for
$k^{-1}\mathrel{\rlap{\lower4pt\hbox{\hskip1pt$\sim$}}
    \raise1pt\hbox{$>$}}10^{3}\,\mbox{Mpc}$) \cite{peeb,lily,star}.
\bibitem{sawo} R. K. Sachs and A. M. Wolfe, Astrophys. J. {\bf 147}, 73 (1967).
\bibitem{grze}
 L. P. Grishchuk and Ya. B. Zel'dovich, Sov. Astron.
{\bf 22}, 125 (1978).
\bibitem{rsxd}
H. Russ, M. H. Soffel, C. Xu and P. K. S. Dunsby, preprint.
\end{references}
\end{document}